\documentclass[iop,apj,tighten]{emulateapj}
\usepackage{apjfonts}

\usepackage{bm}
\usepackage{amsmath,amstext}
\usepackage[breaklinks,colorlinks,citecolor=blue,linkcolor=magenta]{hyperref}
\usepackage[all]{hypcap} 

\shortauthors{Morningstar et al.}
\shorttitle{ALMA Lens Modeling with CNNs}

\usepackage{color}

\newcommand{\CN}{\ensuremath{{\bf C}_N}}

\begin{document}

\title{Analyzing interferometric observations of strong gravitational lenses \\ with recurrent and convolutional neural networks}
\author{Warren R. Morningstar\altaffilmark{1,2}}
\author{Yashar D. Hezaveh\altaffilmark{1}}
\author{Laurence Perreault Levasseur\altaffilmark{1}}
\author{Roger D. Blandford\altaffilmark{1,2}}
\author{Philip J. Marshall\altaffilmark{2}}
\author{Patrick Putzky\altaffilmark{3}}
\author{Risa H. Wechsler\altaffilmark{1,2}}
\altaffiltext{1}{Kavli Institute for Particle Astrophysics and Cosmology and Department of Physics, Stanford University, 452 Lomita Mall, Stanford, CA 94305-4085, USA}
\altaffiltext{2}{Kavli Institute for Particle Astrophysics and Cosmology, SLAC National Accelerator Laboratory, Menlo Park, CA 94025, USA}
\altaffiltext{3}{Informatics Institute, University of Amsterdam}

\begin{abstract}
We use convolutional neural networks (CNNs) and recurrent neural networks (RNNs) to estimate the parameters of strong gravitational lenses from interferometric observations. We explore multiple strategies, including training a feed-forward CNN on dirty images, and find that neural networks can simultaneously adapt to dirty images generated from vastly different $uv$-coverages. We find that the best results are obtained when the effects of the dirty beam are first removed from the images with a deconvolution performed with an RNN-based structure before estimating the parameters. For this purpose, we use the recurrent inference machine (RIM) introduced in \citet{Putzky:17}. This provides a fast and automated alternative to the traditional CLEAN algorithm. We obtain the uncertainties of the estimated parameters using variational inference with Bernoulli distributions. We test the performance of the networks with a simulated test dataset as well as with five ALMA observations of strong lenses. For the observed ALMA data we compare our estimates with values obtained from a maximum-likelihood lens modeling method which operates in the visibility space and find consistent results. We show that we can estimate the lensing parameters with high accuracy using a combination of an RNN structure performing image deconvolution and a CNN performing lensing analysis, with uncertainties less than a factor of two higher than those achieved with maximum-likelihood methods. Including the deconvolution procedure performed by RIM, a single evaluation can be done in about a second on a single GPU, providing a more than six orders of magnitude increase in analysis speed while using about eight orders of magnitude less computational resources compared to maximum-likelihood lens modeling in the $uv$-plane.  We conclude that this is a promising method for the analysis of $mm$ and $cm$ interferometric data from current facilities (e.g., ALMA, JVLA) and future large interferometric observatories (e.g., SKA), where an analysis in the $uv$-plane could be difficult or unfeasible.  
\end{abstract}

 
 \section{Introduction}
Strong gravitational lensing provides a unique opportunity to investigate many subjects, including the distribution of matter in lensing galaxies \citep[e.g., ][]{Treu:04}, the properties of distant galaxies by magnifying their images \citep[e.g., ][]{Jones:10}, and the expansion rate of the universe \citep[e.g., ][]{Suyu:14}. 
Over the past few years, the Atacama Large Millimeter/sub-Millimeter Array (ALMA) has proven to be a unique, powerful tool for imaging sub-millimeter-bright gravitational lenses. ALMA observations of this population of lenses, which were discovered in wide area surveys \citep{vieira:10,negrello:10,vieira:13,hezaveh:13b}, are now allowing significant advances in our understanding of star formation in some of the most active high redshift galaxies \citep[e.g., ][]{Marrone:18}, as well as detailed matter distribution in the foreground structures \citep{Wong:15,Hezaveh:16,Inoue:16,Wong:17}. These studies owe their success to the high sensitivity of these observations and the high resolutions obtained with long baseline interferometry.

The exploitation of strong lensing systems for these studies, however, requires a knowledge of the lensing distortions, traditionally obtained using maximum-likelihood (or \textit{a posteriori}) lens modeling, a procedure in which the posterior of the parameters of a simulated model given the data is maximized. In these methods the values of a set of parameters which describe the true morphology of the background source and the matter distribution in the foreground lens are explored in order to produce a simulated model that best matches the observations. 

Generally, the analysis of lenses with maximum likelihood methods is both slow and technically involved. For example, accurate modeling of optical data requires several data preparation steps, including point spread function (PSF) modeling, subtraction of the lens light, and sophisticated modeling codes. The analysis of interferometric data is even more challenging due to the incomplete sampling of the Fourier space ($uv$-space), where data is measured. The most accurate methods fit the data directly in the $uv$-space. However, due to the large number of the measured visibilities and the large number of lensing parameters, these methods require extremely expensive computations \citep[e.g., see ][]{Hezaveh:16}.

Even with a state-of-the-art pipeline, finding the most probable parameters is a lengthy and resource-intensive process, as it involves using optimizers, requiring a large number of computationally expensive evaluations in the complex, multidimensional space of parameters. Depending on the initial conditions given to these optimizers, they can frequently spend extended periods of time exploring sub-optimal local minima, demanding active human involvement and supervision to expedite convergence to the global solution. In addition, estimating the parameter uncertainties are typically performed with Markov Chain Monte Carlo (MCMC) methods, requiring a large number of likelihood evaluations to converge and to fully sample the parameter space.

Recently, \cite{Hezaveh:17} and \cite{Perreault:17} showed that deep convolutional neural networks could estimate the parameters of strong lenses along with their uncertainties for optical data in an extremely fast and automated manner. These methods construct a direct map from the observed data to the lens parameters using a \textit{training} set and as such do not require the production of simulated models for the analysis of new data. 

Convolutional neural networks \citep{LeCun:89} are a class of deep learning methods that process images through a series of convolutional layers. In each layer, the images from the previous layer are convolved with a number of filters (network weights) and processed with a nonlinear activation function to produce a \emph{feature} map. Typically, after a large number of convolutional layers the feature maps are unraveled and fed into a series of fully connected layers. The activations of the last fully connected layer are then interpreted as the predictions of the network for values of interest. The values of the convolutional filters determine the specific mapping between the input and output data. These values are determined in a process called training, where a set of training data, with known correct input-output pairs (labeled data), are presented to the networks. The values of the network weights are then adjusted to allow the networks to find a successful mapping between the input-output pairs for the training data. In practice, this is done by optimizing a cost function. Since the value of the cost function depends on the networks weights, by calculating its gradient with respect to these weights one could find the weights which optimize the cost function. These gradients are generally calculated using back-propagation.

\begin{figure*}[p]
\includegraphics[width=\hsize]{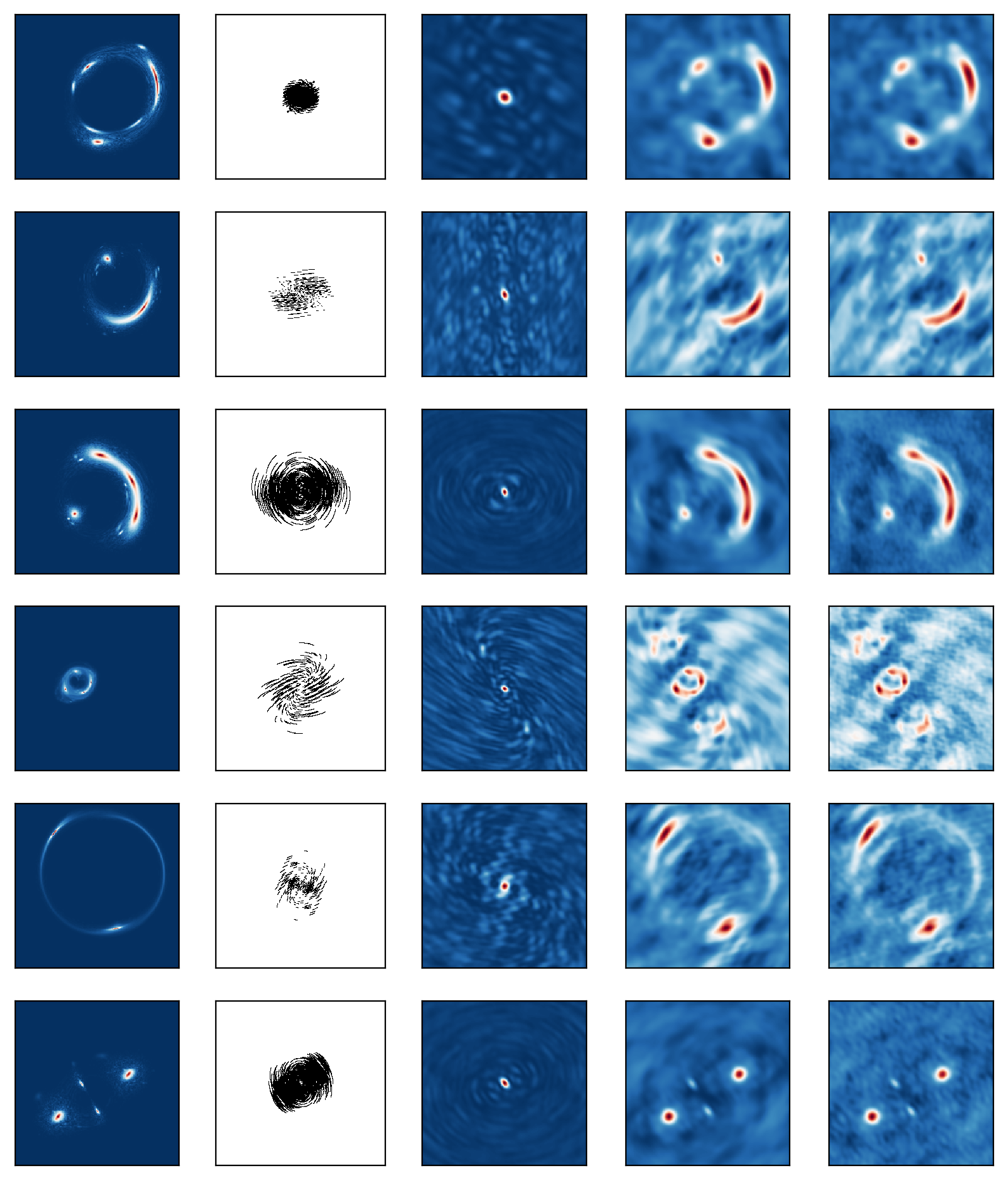}
\caption{Examples of test image simulations. The leftmost column shows the true sky emission created using ray-tracing simulations. Succeeding columns show the randomly produced $uv$-coverages of the observations, the resulting dirty beams, the dirty images, and the noisy dirty images. Qualitatively, the images in column 5 appear significantly different from each other due to the convolution with different beams with significant side lobes.}\label{fig:trainingset}
\end{figure*}

Typically, neural networks are used for point estimation of the outputs of interest. However, it is also possible to obtain the uncertainties of their predictions. An approximate uncertainty estimate could be obtained by training networks to \emph{predict} their own uncertainties. In practice, this can be done by training networks to predict the parameters of an approximating parameter probability distribution. For example, if a Gaussian distribution is used, the networks are required to predict the mean and the variance of the probability distribution of the output parameters. However, since networks can make errors in their own uncertainty estimates, it is essential to marginalize over these network-dependent sources of errors. This can be done using Bayesian neural networks \citep{Neal:96,MacKay:92}. In Bayesian neural networks, instead of fixed deterministic values, the networks weights are defined by probability distributions. In this way, the probability of the weights represents the probability of a certain output. By marginalizing over these distributions then we can marginalize over the network-dependent sources of errors. By using new approximating methods like variational inference \citep{Gal:16}, \cite{Perreault:17} showed that deep convolutional networks could accurately estimate the uncertainties of lens parameters.

In this paper, we expand these studies to the analysis of interferometric observations of gravitational lenses. We explore the use of feed-forward deep convolutional neural networks for estimating the lens parameters from dirty images as well as images produced from deconvolving the effects of the primary beam using a recurrent neural network structure. We do this using the recurrent inference machine \citep{Putzky:17}. We obtain the uncertainties of our predictions using the methodology outlined in \cite{Perreault:17}.  In Section~\ref{sec:methods} we describe the methods and the models that have been explored.  In Section~\ref{sec:results} we report our results, by testing the performance of the networks on simulated and real ALMA data. Finally, in section~\ref{sec:discussion} we discuss the results and the future directions for this work.

\section{methods}\label{sec:methods}
In this section we describe the training data, the architecture of the networks, the strategies explored for training the networks, and the performance tests.

We explore the estimation of lensing parameters from dirty images produced from interferometric observations. Dirty images are obtained by a simple inverse Fourier transform of the visibilities and in essence hold the same information content as the visibilities. However, due to the incomplete sampling of the Fourier space, the resulting beam (the point spread function) is not localized and includes numerous side lobes. This results in the smearing of the signal and the noise over the dirty images, causing correlated structures across the images. In principle, an accurate analysis of interferometric data from dirty images is possible, however, this requires convolving the sky models with the dirty beam and including a dense, correlated noise covariance matrix in the computations of the likelihood functions. This is a complex and computationally expensive task, so many methods in the past have resorted to CLEAN images. In this case, a non-linear algorithm is used to remove the long-range correlations caused by the side-lobes of the dirty beam, resulting in images that resemble CCD images. The CLEAN beam and uncorrelated noise are then used to do the analysis of the data, similar to the analysis of CCD images. These deconvolution methods, in essence, \emph{predict} the missing Fourier modes of the image that are not sampled during observations, by assuming certain priors on the spatial structures of the observed targets. The traditional CLEAN algorithm \citep{Hogbom:74}, for example, assumes that the targets are composed of a collection of distinct point sources. Therefore it provides good results for point sources, but its performance is degraded for targets with extended structures. In all cases, these methods are approximate, complex, non-linear and irreversible procedures that can introduce unknown artifacts in the images, causing biases in the inferred parameters, which cannot be trivially quantified and corrected for. This is why some works have chosen to directly model the visibilities in a space where noise is simple, Gaussian, and uncorrelated \citep{hezaveh:13b,bussmann:13,rybak:2015a, Hezaveh:16}.

In this work we explore the analysis of dirty images using convolutional neural networks. Instead of using a likelihood function, these networks find a mapping between their input data and the outputs of interest through training. They have been shown to be able to adapt to complex structures in their inputs, such as highly correlated images, or correlated noise. It is therefore possible that they can learn to perform an accurate analysis from the highly correlated dirty images. However, since different observations sample different modes in the $uv$-space (due to numerous factors, e.g., observation length, antenna positions, etc.), the resulting dirty images for different observations of the same source can have sharply different appearances and correlations. Here we examine to what extent convolutional neural networks can ignore these structures in a general way. We also explore the prediction of the true sky emission (deconvolved images) directly from the visibilities using the recurrent inference machine prior to estimating the parameters.

\subsection{Training Set}\label{sec:trainingset}

We use the simulated, strongly lensed images of background galaxies described in \cite{Hezaveh:17} and \cite{Perreault:17} to produce a sample of dirty images for training. Here we briefly summarize the procedure used to simulate these images. Real images of local and high-redshift galaxies from the GalaxyZoo and GREAT03 datasets are lensed with an Singular Isothermal Ellipsoid \citep[SIE, ][]{Kormann:94} profile plus external shear. Here, we use $200,000$ unique lensed images. In each case, we ensure that the images have a minimum flux magnification of 3, and that the entire flux is contained within the images. The lens parameters are chosen from uniform random distributions ranging from 0.1 to 3.0 arcseconds in Einstein radius ($\theta_{E}$), 0 to 1 in ellipticity, -0.5 to 0.5 in x and y position, and -0.3 to 0.3 in both components of the external shear.
To avoid the degeneracy of the orientation angle by $\pi$, instead of estimating an angle for ellipticity and external shear, we predict the real and imaginary components of complex ellipticity and shear ($\epsilon_{x}$, $\epsilon_{y}$, $\gamma_{x}$, $\gamma_{y}$). In addition to the seven parameters of the SIE and external shear model, the networks also estimate the total lensing flux magnification ($\mu_{F}$).

\begin{figure*}[tb]
\includegraphics[width=\hsize]{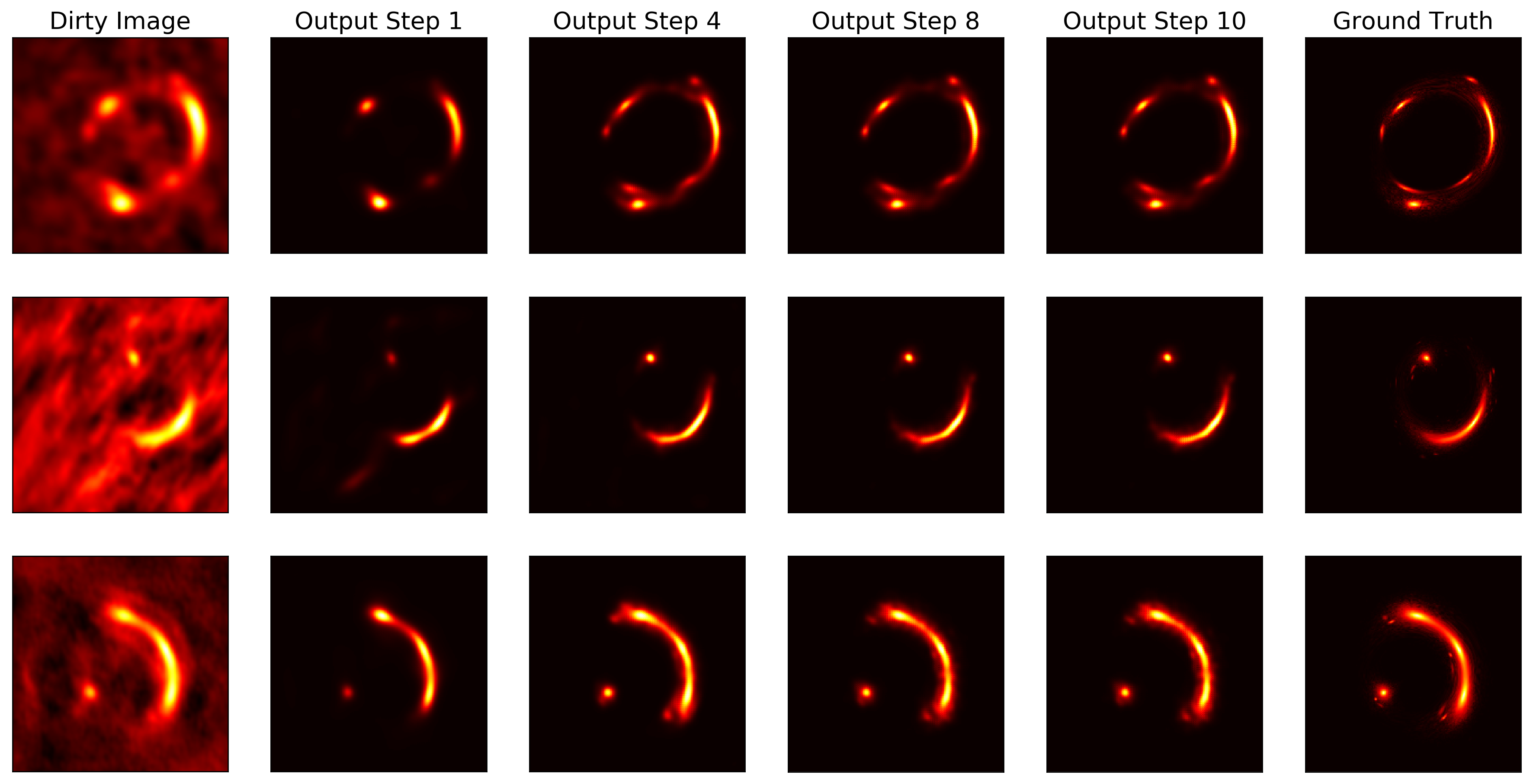}
\caption{Three example image reconstructions through time from the recurrent inference machine. The dirty image (left column) is the inverse Fourier transform of the visibilities and contains significant correlated structures. Through successive passes to the RIM, the underlying signal is iteratively reconstructed. The second column from the right shows the output of the RIM after 10 iterations, which is then fed to the parameter estimation network. The right column shows the ground truth images for comparison.}\label{fig:RIM}
\end{figure*}

These images are then used to produce dirty images resulting from randomly generated $uv$-coverages. We first randomly choose the parameters of the observations: observation start time, duration, source position, the number of antennas used in the observation, and the positions of those antennas.  Assuming an integration time of 60 seconds we then simulate the $uv$-coverage of the observations.  This is a sufficient condition to ensure that no baseline in our training set moves more than a single antenna diameter between separate time steps, and thus any shorter integration time would only add redundant information. In principle, one should then compute the direct Fourier transform of the sky image to predict these visibilities. To produce a dirty image in a fast way, however, these visibilities are typically binned on a regular grid to allow the use of a Fast Fourier Transform (FFT) algorithm. In this work, we introduce an approximation in our production of these dirty images, by first binning our ungridded $uv$-points, and then predicting the visibilities on this regularly-spaced $uv$-grid, using FFT, and then again using FFT to perform the inverse Fourier transform. In practice, this means that we first apply FFT to the sky images, then apply a weighting to the resulting Fourier maps. We simply set the un-sampled modes to zero and scale the measured modes by a weight depending on the number of measured visibilities in that bin. In all cases, we assume similar noise variance in each un-binned visibility. Therefore the noise for a binned visibility is scaled by the square-root of the number of the visibilities contained in it. Random, uncorrelated Gaussian noise is then added to the real and imaginary components of the visibilities. The noise rms is chosen from a uniform random distribution such that the images have peak signal to noise ratios between 10 and 1000. We then obtain the dirty images by taking the inverse Fourier transform of this map using FFT.  To produce a range of effective resolutions during training and to accommodate the variety of $uv$-coverages, we used maximum baselines ranging from $125{\rm k}\lambda$ to $2.2{\rm M}\lambda$, resulting in effective resolutions ranging from $\sim 0.09$ to $1.6$ arcseconds.  This corresponds to a maximum baselines of $400m$ to $2km$ at $350$ GHz. We use natural weighting for producing the dirty images. Figure \ref{fig:trainingset} shows a few examples from the test dataset. For each example, the true sky, the $uv$-coverage, the dirty beam, the dirty image, and the noisy dirty image are presented.

We use stochastic gradient descent to train the networks. At each iteration, we optimize the cost function using a mini-batch of 50 dirty images. Each dirty image is produced from a different, randomly chosen $uv$-coverage and noise realization. Since these observational effects are generated randomly during training, the networks never encounter the same $uv$-coverage or noise realization more than once during training, reducing the risk of overfitting. 

We follow the method described in \cite{Perreault:17} to train the networks to estimate their uncertainties. We optimize a cost function given by the gaussian log-likelihood of the ground truth given the predicted mean and uncertainty from the network
\begin{equation}\label{eq:loss}
\mathcal{L} = \frac{1}{n}\sum_{i=0}^{n}{(y_{i}-\hat{y}_{i})^{2}\exp{\left(-s_{i}\right)}+s_{i}} \, ,
\end{equation}
where $y_{i}$ is the true value of the $i$-th parameter, and $\hat{y_{i}}$ is the value of that parameter predicted by the network.  In this expression $s_{i}=\log{\sigma_{i}^{2}}$, where $\sigma_{i}^2$ is the variance of the predicted Gaussian distribution representing the network uncertainty for the $i$-th parameter. The values of $\sigma_i$ are never provided to the networks (unlabeled outputs) but are implicitly learned by optimizing the cost function. The second term in equation~\ref{eq:loss} ensures that large values of $\sigma_i$ are penalized, while the first term discriminates against small values. This ensures that in the absence of network errors, for a truly Gaussian parameter estimation the predicted $\sigma$ is the rms of the uncertainties. 

To marginalize over network-dependent sources of errors, we use variational inference with Bernoulli distributions for the network weights, using dropout layers before every weight layer. At test time, we perform Monte Carlo dropout to marginalize over these distributions: we feed the same input multiple times to the network and collect the predictions. We then add the predicted uncertainties given by $\sigma_i$ to these samples. The resulting distributions represent the probability distributions of the output parameters.

\capstartfalse 
\begin{deluxetable*}{lllllllll}
\tablewidth{0pt}
\caption{Table}
\tablecolumns{9}
\tablecaption{Model parameters and uncertainties}
\tablehead{
\colhead{Parameter} & \colhead{$\theta_{E}$ } & \colhead{$\epsilon_{x}$} & \colhead{$\epsilon_{y}$} &  \colhead{$x$} & \colhead{$y$} &  \colhead{$\gamma_{x}$} & \colhead{$\gamma_{y}$} & $\mu_{F}$ \\
\colhead{Parameter Description} & \colhead{Einstein Radius} & \colhead{Ellipticity$_{x}$} & \colhead{Ellipticity$_{y}$} & \colhead{Position} & \colhead{Position} & \colhead{Shear$_{x}$} & \colhead{Shear$_{y}$} & \colhead{Flux magnification}}

\hline \hline
    Parameter & $\theta_{E}$ & $\epsilon_{x}$ &  $\epsilon_{y}$ & $x$ & $y$ & $\gamma_{x}$ & $\gamma_{y}$ & $\mu_{F}$ \\
Parameter Description & Einstein Radius & Ellipticity$_x$  & Ellipticity$_y$  & Position  & Position  & Shear$_x$  & Shear$_y$  & Flux magnification 
\startdata
Model 1 & 0.05 & 0.11 & 0.10 & 0.06 & 0.06 & 0.04 & 0.04 & 1.12 \\
Model 2 & 0.05 & 0.11 & 0.10 & 0.06 & 0.06 & 0.04 & 0.04 & 1.12 \\
Model 3 & 0.06 & 0.12 & 0.11 & 0.06 & 0.06 & 0.04 & 0.04 & 1.28 \\
Model 4 & 0.03 & 0.08 & 0.07 & 0.04 & 0.04 & 0.02 & 0.02 & 0.80 \\
Model 4 bias & $2\times10^{-4}$ & $4\times10^{-3}$ & $1\times10^{-3}$  & $-1\times10^{-3}$ & $-5\times10^{-3}$ & $1\times10^{-3}$ & $2\times10^{-3}$ & 0.3 
\enddata
\tablecomments{Median root-mean-squared uncertainties of network estimated parameters produced by sampling the network predictions on a simulated test set.  We also show the bias on each parameter found using Model 4.  This bias is always substantially smaller than the uncertainty.}\label{tab:precision}
\end{deluxetable*}
\capstarttrue

\begin{figure*}[tb]
\includegraphics[width=\hsize]{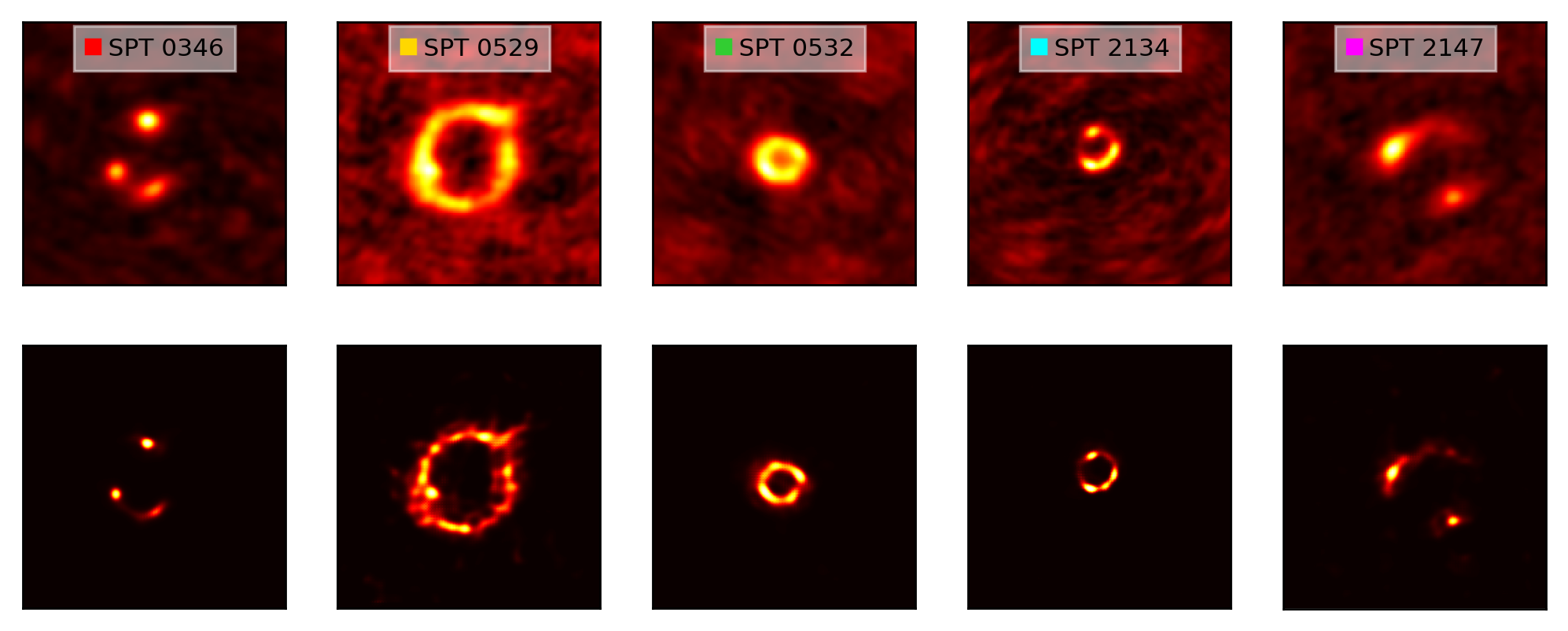}
\caption{ALMA observations of gravitational lenses performed during Cycle 2.  The top panels show the dirty images, created via an inverse Fourier transform of the visibilities.  The bottom panel shows the output images from the RIM.  The colored squares correspond to the plotting symbols shown in Figure~\ref{fig:params}.}\label{fig:Cycle2}
\end{figure*}

\subsection{Training strategies}
As is shown in Figure~\ref{fig:trainingset}, different $uv$-coverages of different observations result in significant differences in the appearance of the images. To test the ability of CNNs to adapt to such variable long-range correlations, we have performed tests with four different models. These models are as follows:

\textbf{Model 1:} We first consider training a network specifically optimized to estimate the lensing parameters for a particular ALMA observation. Therefore, we produce training dirty images that result from sampling the specific $uv$-coverage of the observation under consideration. To do this, we calculate a dirty beam using a direct Fourier transform of the $uv$-coordinates and convolve the true sky emission with this beam to produce a dirty image. We also produce noise maps resulting from random, Gaussian, uncorrelated noise in the measured visibilities and randomly add them with different overall scaling to the dirty images. The use of direct Fourier transform ensures that no artifacts from visibility binning are introduced.

\textbf{Model 2:}  We then consider training a network to estimate the lensing parameters from multiple distinct ALMA observations with different $uv$-coverages simultaneously.  The generation of training examples for this network is the same as Model 1, but instead of using a single $uv$-configuration to generate the dirty images, we use noise maps and dirty beams from five separate $uv$-configurations given by ALMA Cycle 2 observations of five strong gravitational lenses.  The $uv$-coverages used for a given training example is selected randomly from the five choices with equal probability.  

\textbf{Model 3:}  Next we consider training a network to estimate the lensing parameters for \emph{any} ALMA observation with an arbitrary $uv$-coverage.  To do this, during training, we randomly generate dirty images from different $uv$-configurations as described in section~\ref{sec:trainingset}.

\textbf{Model 4:}  Finally, we consider using a network to remove the effects of the dirty beam (deconvolution) prior to parameter estimation.  To do this, we use the framework of a recurrent inference machine (RIM) developed in \cite{Putzky:17} to reconstruct the image. This network performs an iterative procedure, using recurrent convolutional neural networks, to iteratively solve the linear equation $\textbf{y}=\textbf{A}\, \textbf{x}+\textbf{n}$ for $\textbf{x}$, where $\textbf{y}$ is a vector of measurements, $\textbf{A}$ is a corruption matrix, and $\textbf{n}$ is an additive noise vector. In the present application, $\textbf{x}$ is the true sky emission, $\textbf{y}$ is the observed visibilities, $\textbf{A}$ is a Fourier transform matrix, and $\textbf{n}$ is a vector of additive uncorrelated Gaussian noise.

We refer the reader to \cite{Putzky:17} for further details, but in principle, the RIM architecture solves this equation by using the gradient of the log-likelihood of $\textbf{y}$ given $\textbf{x}$ with respect to $\textbf{x}$, evaluated at the current estimate of $\textbf{x}$, in a fashion analogous to the Newton's method for optimization. Here we compute the visibilities of the model image with FFT and calculate the log-likelihood in the visibility space:
\begin{equation}\label{eq:log-likelihood}
\mathcal{L}(I) =[V_{obs}-\mathcal{F}(I)]^{T}\CN^{-1}[V_{obs}-\mathcal{F}(I)] \, ,
\end{equation}
where $\mathcal{F}$ denotes the operation of predicting the visibilities from the sky emission, $I$ is the predicted image, $\CN$ is the noise covariance matrix, and $V_{obs}$ are the observed visibilities.
We then take the gradient of this likelihood with respect to the image pixels.  
To reduce the error introduced by using FFTs instead of direct Fourier transforms and the periodic boundary conditions of FFT, we pad the input images to obtain higher resolution results in the visibility space.  The errors on the resulting dirty image pixel values produced by this gridding are less than $0.1\%$ of the peak image value and more than 100 times smaller than the typical noise rms. The resulting deconvolved images are then fed to a separate, feed-forward convolutional neural network that estimates the lensing parameters. 

Because the purpose of the RIM is to produce deconvolved images, we use the mean-squared error over all the pixels and over all time steps as a cost function.  By optimizing over all time steps, we allow gradients to propagate back through the network more easily.  This facilitates more efficient training, especially in the early stages of the optimization.  We train the RIM separately from the CNN.  Once it is adequately trained, we then fix all of its parameters, and use its output to train the feed-forward CNN.  Similar to Model 3, Model 4 is trained using randomly generated $uv$-configurations as described in Section~\ref{sec:trainingset}.

All models were implemented in TensorFlow \citep{Tensorflow}.  For all models, for estimating the lensing parameters, we use the architecture of AlexNet \citep{Krizhevsky:12}, which is of relatively modest size (16 million parameters) and has been shown to perform well for lens analysis \citep{Hezaveh:17}. The last layer of the network predicts 16 values, corresponding to the eight parameters of interest and their marginalized uncertainties.

\begin{figure*}[tb]
\includegraphics[width=\hsize]{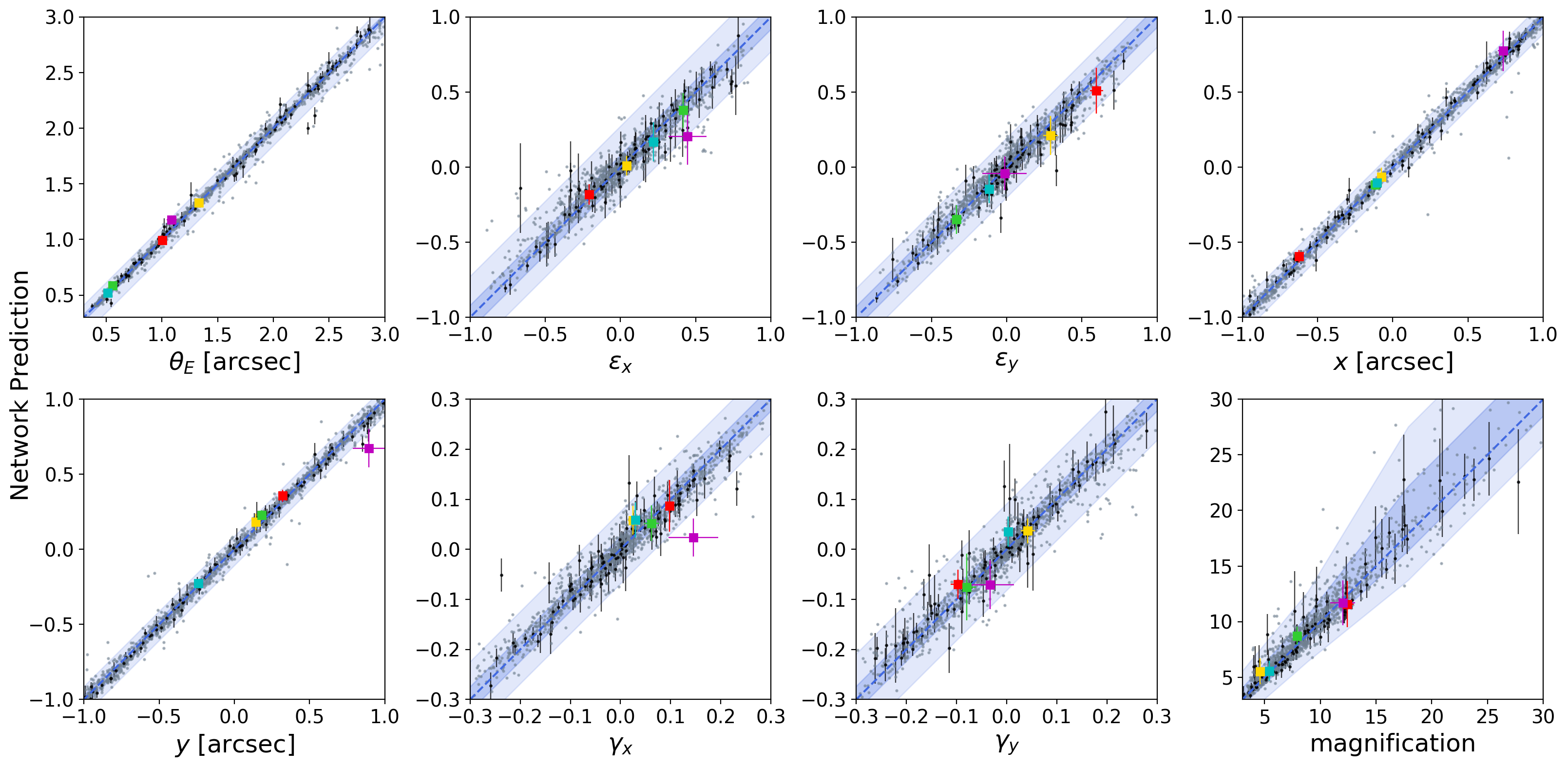}
\caption{Comparison of the true values of the lens parameters ($x$-axis) with  their estimated values ($y$-axis) using Model 4. The gray points show the mean of the predicted values for each example. For a small subset of the examples, the $1$-$\sigma$ uncertainties of the predictions are also shown with error bars. The light and dark blue bands show the intervals containing $68$ and $95\%$ of these mean values from the true values.  The dashed line indicates the correct prediction ($y=x$ line). The colored points and error bars show the predicted parameter values for the five ALMA Cycle 2 observations of strongly lensed sources, where the colors correspond to the colors in Figure~\ref{fig:Cycle2}. For these sources, the values on the $x$-axis and their uncertainties are obtained by a MAP modeling of the observations.}\label{fig:params}
\end{figure*}

We train our network using the Adam optimizer \citep{Kingma:14}.  This algorithm uses an exponentially weighted average of the past gradients as a "momentum" and updates the network parameters using the momentum rather than the gradient itself.  This causes individual training steps to be smoothed out, reducing the stochasticity of the optimization process and allowing for more efficient minimization of the cost function. We use a learning rate schedule, starting at $2\times10^{-5}$ for 200,000 training steps, and subsequently stepping down by a factor of two every 50,000 training steps.  

The dropout rate is tuned as a hyperparameter that can be empirically adjusted to calibrate the predicted uncertainties, following the procedure described in \cite{Perreault:17}.  We train an ensemble of networks, each with a different dropout rate (ranging from 0.5 to 0.0, or equivalently a keep rate of 0.5 to 1.0).  We then calculate the coverage probabilities of the resulting uncertainties for each trained network with a validation set. We then select a keep rate that results in coverage probabilities equal to to the 1-, 2-, and 3-$\sigma$ confidence levels (i.e. the 68\% confidence interval should have a coverage probability of 68\% by construction). We find that the coverage probabilities are best matched with a keep rate of 99\%. We therefore adopt a dropout rate of 1\% for the rest of this work.

\subsection{Performance tests}

To quantify the performance of the networks, we test their predictions on a separate test set of $\sim1400$ simulated lensed images. In all cases, while the lensed images are different from the images in the training set, they undergo the same data processing as the training data prior to being fed to the network.  This means that to test models 1 and 2 we use the same dirty beams that were used in training, and for models 3 and 4 we produce dirty images from randomly generated $uv$-coverages.

In addition to simulated data, we also test the performance of the networks on real ALMA observations of gravitational lenses. We use ALMA observations of five gravitational lenses observed during ALMA Cycle 2 (2013.1.00880.S, PI: Hezaveh). These observations are approximately 40 minutes in duration and were taken using 37 ALMA antennae with a maximum baseline of $1500 \textrm{m}$ at a frequency of 145 GHz (ALMA Band 4). For comparison, we also model these targets with a maximum a posteriori lens modeling pipeline that treats the background source as a vector of pixels \citep{Hezaveh:16}.

\begin{figure*}[tb]
\includegraphics[trim= 0 0 0 0, clip, width=1.0\hsize]{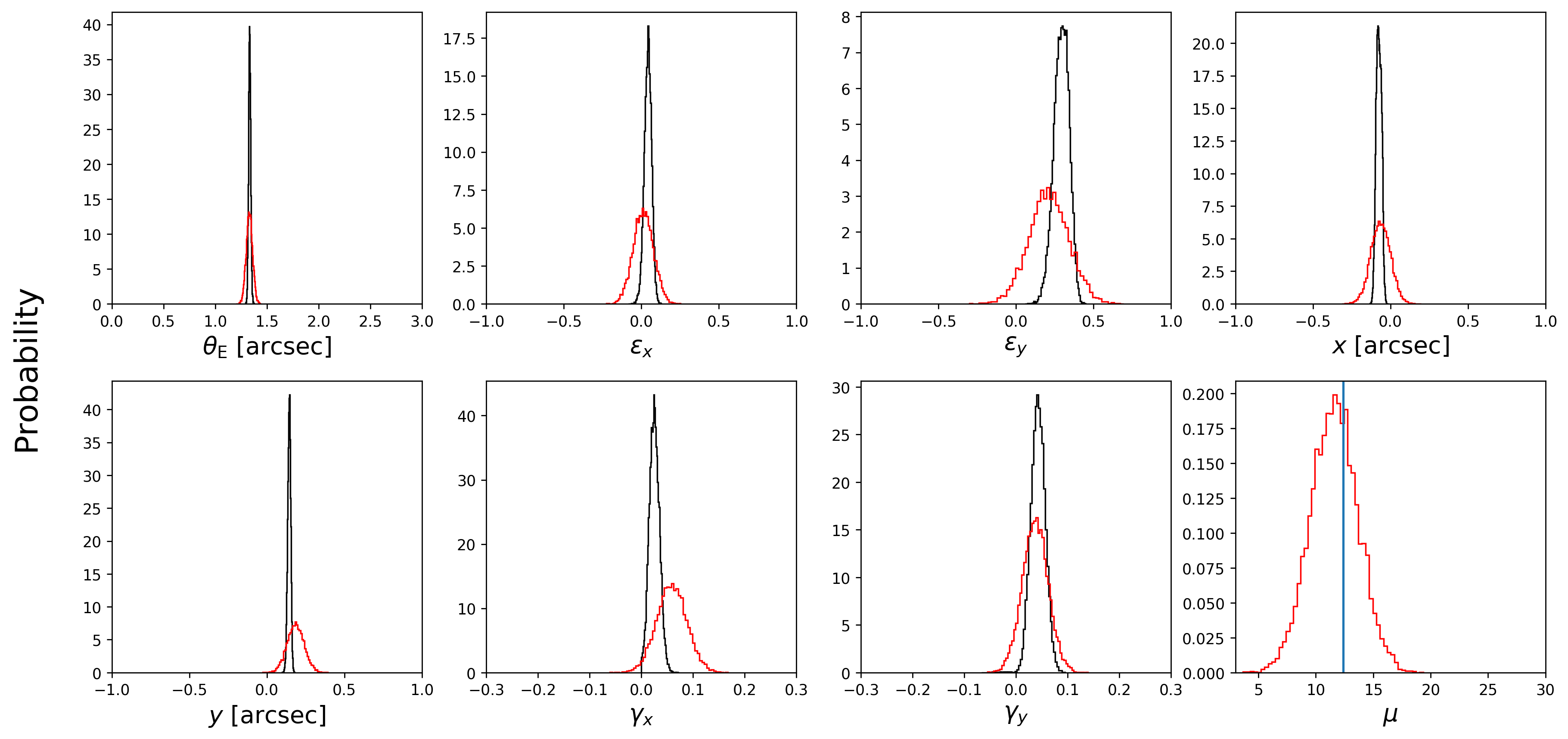}
\caption{Comparison of the predictions of Model 4 (red) to the predictions of a maximum a posteriori pixellated lens modeling pipeline (black) for gravitational lensing system SPT 0529.  The axis limits indicate the range of each parameter used during training. All parameters are accurately recovered by the network but with larger uncertainties.  These results are similar for the other four ALMA observations.}\label{fig:distribution}
\end{figure*}

\section{results}\label{sec:results}

Table~\ref{tab:precision} shows the median uncertainty produced by the four training strategies. Models 1 and 2 perform similarly well but slightly better than model 3. This is to be expected, given that the network in Model 3 has to adapt to any arbitrary $uv$-coverage, while Models 1 and 2 have been specialized to one or a small subset of relatively similar $uv$-coverages.

Model 4 outperforms all the other models. We attribute this to the ability of the RIM to remove the effects of the dirty beam in a general way, producing  consistent images for the parameter estimation network. As a measure of the bias of the parameter estimation network, we calculate the mean of the difference between the predicted and true values of the lens model parameters and report them in Table~\ref{tab:precision}.  Figure~\ref{fig:RIM} shows example reconstructions of the sky images using the RIM. For each example, the original dirty image and the output of the RIM at different iterations are shown and compared to the true sky emission.  Compared to the dirty images, which have substantially different properties due to the beam and correlated noise, the RIM outputs appear qualitatively similar except for the differences due to lensing.  Figure~\ref{fig:Cycle2} shows the output of the final step of the RIM reconstruction applied to ALMA observations.

Figure~\ref{fig:params} shows the predicted parameter values for Model 4 for the simulated data against the ground truth values. The mean values of the predicted parameters for all the 1400 examples are shown with small dots. The light and dark blue bands show the intervals containing $68$, and $95\%$ of these mean values. For a small subset of the examples, the $1$-$\sigma$ uncertainties of the predictions are also shown with error bars. The square points show the predicted parameter values and their uncertainties for the five ALMA Cycle 2 observations of strongly lensed sources. For these sources, the values on the $x$-axis and their uncertainties are obtained by a \textit{maximum a posteriori} (MAP) modeling of the observations.

For most examples the predictions are an excellent approximation to the true values. The estimates with the largest errors also have large uncertainties associated with them, such that the overall coverage probabilities are equal to their confidence limits for which they are calculated.  By examining the images in the test set with the largest errors, we find that they consist of instances of doubly-imaged lenses with no extended arcs or images in naked-cusp configuration. It is well known that these configurations typically have less constraining power compared to configurations with more extended arcs and counter images and result in larger uncertainties with MAP modeling method as well \citep{Nightingale:18}.

Figure~\ref{fig:distribution} shows the probability distributions for each parameter obtained with model 4 and MAP modeling for one of the sources from ALMA Cycle 2 data. We find that the neural networks produce distributions that are consistent with MAP models but slightly broader. However they only take of order 1 second to fully sample the network posterior on a single GPU. Compared to over a month on $\sim1000$ CPU cores required to sample the posteriors of these observations with traditional lens modeling methods, this results in more than 6 orders of magnitude speed-up of the analysis. Of course, in reality the gains are even larger due to the time required for finding the global optimum and MCMC convergence.

\section{discussion and conclusion}\label{sec:discussion}

The results presented in the previous section demonstrate that neural networks can accurately estimate lensing model parameters from interferometric observations in an extremely fast manner. The uncertainties obtained using a Recurrent Inference Machine and Convolutional Neural Network (model 4) are typically less than a factor of two higher than the uncertainties obtained from MAP modeling, however, they provide more than six orders of magnitude speed-up in wall clock time while using about eight orders of magnitude less computational resources. 

Although the use of the RIM for deconvolving the dirty beam is somewhat similar to the CLEAN algorithm, it has several advantages. First, the morphology of the images in the training data act as a prior for the reconstructions. This can result in improved performance when the test images have similar properties to those of the training data. Second, once trained, this is a fully automated procedure and does not require any manual adjustments (e.g., mask defining, stopping criteria). Third, as was done in this work, because of the high speed (of order 1 second) and the fully automated nature of the RIM, it is possible to produce large uniform samples from it and train an analysis network with their outputs. This ensures that if there are systematic errors introduced by the RIM, the analysis networks can learn to ignore them and include their effects in their final uncertainties. This is a significant improvement to traditional deconvolution algorithms like CLEAN, where possible artifacts could depend on user-defined settings and can not be tracked or included in the final uncertainties. Fourth, since the deconvolved image is produced by maximizing the likelihood in the visibility space, this results in output images with better fidelity to the original measured visibilities compared to the CLEAN algorithm. 

More generally, the speed of the predictions and the calibration of the uncertainties ensures that the coverage probabilities calculated over a large set of examples are equal to the confidence limits for which they are calculated. In other words, this means that these uncertainties include the contributions of systematic errors. It is well known that MAP lens modeling can sometimes result in biased parameter recovery due to numerous effects including the choice of source parameterization. It is therefore likely that the parameters recovered with these networks can be more accurate than those predicted with MAP methods.

Perhaps the most important element in this analysis is the design of the training data. In particular, since we have used simulated data to train a network, which is then used for the interpretation of real data, special care should be given to understanding the structure and the statistical properties of real data and to define a training set which encompasses the variations of all possible effects in the real data. In this work, we used a few approximations to produce the training set (e.g., gridding the $uv$-coordinates prior to predicting the visibilities). For the purpose of the demonstration of the method in this paper, these approximations seem justified, given that the recovered parameters for real ALMA observations of SPT sources are consistent with their values from MAP modeling. However, if these methods are going to be widely used for real data analysis, it is preferable to produce even more realistic training data. In addition, it is possible to use domain adaptation methods \citep[e.g., ][]{Ben-David} to generalize the learning of the networks from simulated examples to real data with different statistical properties.

We explored strategies for the analysis of strong gravitational lenses from interferometric data with neural networks. We found that it is possible to train simple feed-forward convolutional neural networks on dirty images produced from the measured visibilities, however, the best results were obtained when a recurrent neural network-based architecture was first used to remove the effects of the convolution of the sky emission with the dirty beam prior to estimating the parameters using feed-forward models. This method produced estimates with a median precision comparable to MAP modeling (typically less than a factor of two lower), while resulting in orders of magnitude improvement in speed and the use of computational resources. Given the large number of observations expected to be executed by ALMA and other interferometric facilities, these methods can be a crucial tool for the interpretation of future data.

\acknowledgements
We are grateful to Max Welling for discussions about the RIM architecture.  We would also like to thank Dan Marrone, Neal Dalal, Joaquin Vieira, and Gil Holder for reading the manuscript and for their involvement in the data acquisition.  Support for this work was provided by the National Science Foundation through NSF AST-1716527, and by NASA through Hubble Fellowship grant HST-HF2-51358.001-A by the Space Telescope Science Institute, operated by the Association of Universities for Research in Astronomy, Inc., for NASA, under contract NAS 5-26555. Additional support was provided by the U.S. Department of Energy under contract number DE-AC02-76SF00515. Computing for this project was performed on the Sherlock and XStream clusters. We would like to thank Stanford University and the Stanford Research Computing Center for providing computational resources and support that contributed to this work. XStream was supported by the National Science Foundation Major Research Instrumentation program (ACI-1429830). 

This paper makes use of the following ALMA data: ADS/JAO.ALMA\#2013.1.00880.S. 
ALMA is a partnership of ESO (representing its member states), NSF (USA) and 
NINS (Japan), together with NRC (Canada) and NSC and ASIAA (Taiwan), and KASI (Republic of Korea), in cooperation with the Republic of Chile. The Joint ALMA Observatory is operated by ESO, AUI/NRAO and NAOJ. The National Radio Astronomy Observatory is a facility of the National Science Foundation operated under cooperative agreement by Associated Universities, Inc.

\bibliographystyle{yahapj}

\begin{thebibliography}{}
\providecommand\natexlab[1]{#1}
\providecommand\JournalTitle[1]{#1}

\bibitem[{Abadi {et~al.}(2015)Abadi, Agarwal, Barham, Brevdo, Chen, Citro,
  Corrado, Davis, Dean, Devin, Ghemawat, Goodfellow, Harp, Irving, Isard, Jia,
  Jozefowicz, Kaiser, Kudlur, Levenberg, Man\'{e}, Monga, Moore, Murray, Olah,
  Schuster, Shlens, Steiner, Sutskever, Talwar, Tucker, Vanhoucke, Vasudevan,
  Vi\'{e}gas, Vinyals, Warden, Wattenberg, Wicke, Yu, \& Zheng}]{Tensorflow}
Abadi, M., Agarwal, A., Barham, P., {et~al.} 2015, {TensorFlow}: Large-Scale
  Machine Learning on Heterogeneous Systems, software available from
  tensorflow.org

\bibitem[{Ben-David {et~al.}(2007)Ben-David, Blitzer, Crammer, \&
  Pereira}]{Ben-David}
Ben-David, S., Blitzer, J., Crammer, K., \& Pereira, F. 2007,
  \href{http://papers.nips.cc/paper/2983-analysis-of-representations-for-domain-adaptation.pdf}{in
  Advances in Neural Information Processing Systems 19, ed. B.~Sch\"{o}lkopf,
  J.~C. Platt, \& T.~Hoffman} (MIT Press), 137

\bibitem[{{Bussmann} {et~al.}(2013){Bussmann}, {P{\'e}rez-Fournon}, {Amber},
  {Calanog}, {Gurwell}, {Dannerbauer}, {De Bernardis}, {Fu}, {Harris}, {Krips},
  {Lapi}, {Maiolino}, {Omont}, {Riechers}, {Wardlow}, {Baker}, {Birkinshaw},
  {Bock}, {Bourne}, {Clements}, {Cooray}, {De Zotti}, {Dunne}, {Dye}, {Eales},
  {Farrah}, {Gavazzi}, {Gonz{\'a}lez Nuevo}, {Hopwood}, {Ibar}, {Ivison},
  {Laporte}, {Maddox}, {Mart{\'{\i}}nez-Navajas}, {Michalowski}, {Negrello},
  {Oliver}, {Roseboom}, {Scott}, {Serjeant}, {Smith}, {Smith}, {Streblyanska},
  {Valiante}, {van der Werf}, {Verma}, {Vieira}, {Wang}, \&
  {Wilner}}]{bussmann:13}
{Bussmann}, R.~S., {P{\'e}rez-Fournon}, I., {Amber}, S., {et~al.} 2013,
  \href{http://dx.doi.org/10.1088/0004-637X/779/1/25}{\JournalTitle{\apj}, 779,
  25}

\bibitem[{Gal \& Ghahramani(2016)}]{Gal:16}
Gal, Y., \& Ghahramani, Z. 2016, in international conference on machine
  learning, 1050

\bibitem[{{Hezaveh} {et~al.}(2017){Hezaveh}, {Levasseur}, \&
  {Marshall}}]{Hezaveh:17}
{Hezaveh}, Y.~D., {Levasseur}, L.~P., \& {Marshall}, P.~J. 2017,
  \href{http://dx.doi.org/10.1038/nature23463}{\JournalTitle{\nat}, 548, 555}

\bibitem[{{Hezaveh} {et~al.}(2013){Hezaveh}, {Marrone}, {Fassnacht}, {Spilker},
  {Vieira}, {Aguirre}, {Aird}, {Aravena}, {Ashby}, {Bayliss}, {Benson},
  {Bleem}, {Bothwell}, {Brodwin}, {Carlstrom}, {Chang}, {Chapman}, {Crawford},
  {Crites}, {De Breuck}, {de Haan}, {Dobbs}, {Fomalont}, {George}, {Gladders},
  {Gonzalez}, {Greve}, {Halverson}, {High}, {Holder}, {Holzapfel}, {Hoover},
  {Hrubes}, {Husband}, {Hunter}, {Keisler}, {Lee}, {Leitch}, {Lueker},
  {Luong-Van}, {Malkan}, {McIntyre}, {McMahon}, {Mehl}, {Menten}, {Meyer},
  {Mocanu}, {Murphy}, {Natoli}, {Padin}, {Plagge}, {Reichardt}, {Rest}, {Ruel},
  {Ruhl}, {Sharon}, {Schaffer}, {Shaw}, {Shirokoff}, {Stalder}, {Staniszewski},
  {Stark}, {Story}, {Vanderlinde}, {Wei{\ss}}, {Welikala}, \&
  {Williamson}}]{hezaveh:13b}
{Hezaveh}, Y.~D., {Marrone}, D.~P., {Fassnacht}, C.~D., {et~al.} 2013,
  \href{http://dx.doi.org/10.1088/0004-637X/767/2/132}{\JournalTitle{\apj},
  767, 132}

\bibitem[{{Hezaveh} {et~al.}(2016){Hezaveh}, {Dalal}, {Marrone}, {Mao},
  {Morningstar}, {Wen}, {Blandford}, {Carlstrom}, {Fassnacht}, {Holder},
  {Kemball}, {Marshall}, {Murray}, {Perreault Levasseur}, {Vieira}, \&
  {Wechsler}}]{Hezaveh:16}
{Hezaveh}, Y.~D., {Dalal}, N., {Marrone}, D.~P., {et~al.} 2016,
  \href{http://dx.doi.org/10.3847/0004-637X/823/1/37}{\JournalTitle{\apj}, 823,
  37}

\bibitem[{{H{\"o}gbom}(1974)}]{Hogbom:74}
{H{\"o}gbom}, J.~A. 1974, \JournalTitle{\aaps}, 15, 417

\bibitem[{{Inoue} {et~al.}(2016){Inoue}, {Minezaki}, {Matsushita}, \&
  {Chiba}}]{Inoue:16}
{Inoue}, K.~T., {Minezaki}, T., {Matsushita}, S., \& {Chiba}, M. 2016,
  \href{http://dx.doi.org/10.1093/mnras/stw168}{\JournalTitle{\mnras}, 457,
  2936}

\bibitem[{{Jones} {et~al.}(2010){Jones}, {Swinbank}, {Ellis}, {Richard}, \&
  {Stark}}]{Jones:10}
{Jones}, T.~A., {Swinbank}, A.~M., {Ellis}, R.~S., {Richard}, J., \& {Stark},
  D.~P. 2010,
  \href{http://dx.doi.org/10.1111/j.1365-2966.2010.16378.x}{\JournalTitle{\mnras},
  404, 1247}

\bibitem[{Kingma \& Ba(2014)}]{Kingma:14}
Kingma, D.~P., \& Ba, J. 2014,
  \href{http://arxiv.org/abs/1412.6980}{\JournalTitle{CoRR}, abs/1412.6980},
  \href{http://arxiv.org/abs/1412.6980}{{\sffamily arXiv:1412.6980}}

\bibitem[{{Kormann} {et~al.}(1994){Kormann}, {Schneider}, \&
  {Bartelmann}}]{Kormann:94}
{Kormann}, R., {Schneider}, P., \& {Bartelmann}, M. 1994, \JournalTitle{\aap},
  284, 285

\bibitem[{Krizhevsky {et~al.}(2012)Krizhevsky, Sutskever, \&
  Hinton}]{Krizhevsky:12}
Krizhevsky, A., Sutskever, I., \& Hinton, G.~E. 2012,
  \href{http://dl.acm.org/citation.cfm?id=2999134.2999257}{in Proceedings of
  the 25th International Conference on Neural Information Processing Systems -
  Volume 1, NIPS'12} (USA: Curran Associates Inc.), 1097

\bibitem[{LeCun {et~al.}(1989)LeCun, Boser, Denker, Henderson, Howard, Hubbard,
  \& Jackel}]{LeCun:89}
LeCun, Y., Boser, B., Denker, J.~S., {et~al.} 1989, \JournalTitle{Neural
  computation}, 1, 541

\bibitem[{MacKay(1992)}]{MacKay:92}
MacKay, D.~J. 1992, \JournalTitle{Neural computation}, 4, 448

\bibitem[{{Marrone} {et~al.}(2018){Marrone}, {Spilker}, {Hayward}, {Vieira},
  {Aravena}, {Ashby}, {Bayliss}, {B{\'e}thermin}, {Brodwin}, {Bothwell},
  {Carlstrom}, {Chapman}, {Chen}, {Crawford}, {Cunningham}, {De Breuck},
  {Fassnacht}, {Gonzalez}, {Greve}, {Hezaveh}, {Lacaille}, {Litke}, {Lower},
  {Ma}, {Malkan}, {Miller}, {Morningstar}, {Murphy}, {Narayanan}, {Phadke},
  {Rotermund}, {Sreevani}, {Stalder}, {Stark}, {Strandet}, {Tang}, \&
  {Wei{\ss}}}]{Marrone:18}
{Marrone}, D.~P., {Spilker}, J.~S., {Hayward}, C.~C., {et~al.} 2018,
  \href{http://dx.doi.org/10.1038/nature24629}{\JournalTitle{\nat}, 553, 51}

\bibitem[{Neal(1996)}]{Neal:96}
Neal, R.~M. 1996, Bayesian Learning for Neural Networks (Secaucus, NJ, USA:
  Springer-Verlag New York, Inc.)

\bibitem[{{Negrello} {et~al.}(2010){Negrello}, {Hopwood}, {De Zotti}, {Cooray},
  {Verma}, {Bock}, {Frayer}, {Gurwell}, {Omont}, {Neri}, {Dannerbauer},
  {Leeuw}, {Barton}, {Cooke}, {Kim}, {da Cunha}, {Rodighiero}, {Cox},
  {Bonfield}, {Jarvis}, {Serjeant}, {Ivison}, {Dye}, {Aretxaga}, {Hughes},
  {Ibar}, {Bertoldi}, {Valtchanov}, {Eales}, {Dunne}, {Driver}, {Auld},
  {Buttiglione}, {Cava}, {Grady}, {Clements}, {Dariush}, {Fritz}, {Hill},
  {Hornbeck}, {Kelvin}, {Lagache}, {Lopez-Caniego}, {Gonzalez-Nuevo}, {Maddox},
  {Pascale}, {Pohlen}, {Rigby}, {Robotham}, {Simpson}, {Smith}, {Temi},
  {Thompson}, {Woodgate}, {York}, {Aguirre}, {Beelen}, {Blain}, {Baker},
  {Birkinshaw}, {Blundell}, {Bradford}, {Burgarella}, {Danese}, {Dunlop},
  {Fleuren}, {Glenn}, {Harris}, {Kamenetzky}, {Lupu}, {Maddalena}, {Madore},
  {Maloney}, {Matsuhara}, {Michaowski}, {Murphy}, {Naylor}, {Nguyen},
  {Popescu}, {Rawlings}, {Rigopoulou}, {Scott}, {Scott}, {Seibert}, {Smail},
  {Tuffs}, {Vieira}, {van der Werf}, \& {Zmuidzinas}}]{negrello:10}
{Negrello}, M., {Hopwood}, R., {De Zotti}, G., {et~al.} 2010,
  \href{http://dx.doi.org/10.1126/science.1193420}{\JournalTitle{Science}, 330,
  800}

\bibitem[{Nightingale {et~al.}(2018)Nightingale, Dye, \&
  Massey}]{Nightingale:18}
Nightingale, J., Dye, S., \& Massey, R.~J. 2018, \JournalTitle{Monthly Notices
  of the Royal Astronomical Society}, 478, 4738

\bibitem[{{Perreault Levasseur} {et~al.}(2017){Perreault Levasseur}, {Hezaveh},
  \& {Wechsler}}]{Perreault:17}
{Perreault Levasseur}, L., {Hezaveh}, Y.~D., \& {Wechsler}, R.~H. 2017,
  \href{http://dx.doi.org/10.3847/2041-8213/aa9704}{\JournalTitle{\apjl}, 850,
  L7}

\bibitem[{Putzky \& Welling(2017)}]{Putzky:17}
Putzky, P., \& Welling, M. 2017, \JournalTitle{arXiv preprint arXiv:1706.04008}

\bibitem[{{Rybak} {et~al.}(2015){Rybak}, {McKean}, {Vegetti}, {Andreani}, \&
  {White}}]{rybak:2015a}
{Rybak}, M., {McKean}, J.~P., {Vegetti}, S., {Andreani}, P., \& {White},
  S.~D.~M. 2015, \JournalTitle{ArXiv e-prints},
  \href{http://arxiv.org/abs/1503.02025}{{\sffamily arXiv:1503.02025}}

\bibitem[{{Suyu} {et~al.}(2014){Suyu}, {Treu}, {Hilbert}, {Sonnenfeld},
  {Auger}, {Blandford}, {Collett}, {Courbin}, {Fassnacht}, {Koopmans},
  {Marshall}, {Meylan}, {Spiniello}, \& {Tewes}}]{Suyu:14}
{Suyu}, S.~H., {Treu}, T., {Hilbert}, S., {et~al.} 2014,
  \href{http://dx.doi.org/10.1088/2041-8205/788/2/L35}{\JournalTitle{\apjl},
  788, L35}

\bibitem[{{Treu} \& {Koopmans}(2004)}]{Treu:04}
{Treu}, T., \& {Koopmans}, L.~V.~E. 2004,
  \href{http://dx.doi.org/10.1086/422245}{\JournalTitle{\apj}, 611, 739}

\bibitem[{{Vieira} {et~al.}(2010){Vieira}, {Crawford}, {Switzer}, {Ade},
  {Aird}, {Ashby}, {Benson}, {Bleem}, {Brodwin}, {Carlstrom}, {Chang}, {Cho},
  {Crites}, {de Haan}, {Dobbs}, {Everett}, {George}, {Gladders}, {Hall},
  {Halverson}, {High}, {Holder}, {Holzapfel}, {Hrubes}, {Joy}, {Keisler},
  {Knox}, {Lee}, {Leitch}, {Lueker}, {Marrone}, {McIntyre}, {McMahon}, {Mehl},
  {Meyer}, {Mohr}, {Montroy}, {Padin}, {Plagge}, {Pryke}, {Reichardt}, {Ruhl},
  {Schaffer}, {Shaw}, {Shirokoff}, {Spieler}, {Stalder}, {Staniszewski},
  {Stark}, {Vanderlinde}, {Walsh}, {Williamson}, {Yang}, {Zahn}, \&
  {Zenteno}}]{vieira:10}
{Vieira}, J.~D., {Crawford}, T.~M., {Switzer}, E.~R., {et~al.} 2010,
  \href{http://dx.doi.org/10.1088/0004-637X/719/1/763}{\JournalTitle{\apj},
  719, 763}

\bibitem[{{Vieira} {et~al.}(2013){Vieira}, {Marrone}, {Chapman}, {De Breuck},
  {Hezaveh}, {Wei{$\beta$}}, {Aguirre}, {Aird}, {Aravena}, {Ashby}, {Bayliss},
  {Benson}, {Biggs}, {Bleem}, {Bock}, {Bothwell}, {Bradford}, {Brodwin},
  {Carlstrom}, {Chang}, {Crawford}, {Crites}, {de Haan}, {Dobbs}, {Fomalont},
  {Fassnacht}, {George}, {Gladders}, {Gonzalez}, {Greve}, {Gullberg},
  {Halverson}, {High}, {Holder}, {Holzapfel}, {Hoover}, {Hrubes}, {Hunter},
  {Keisler}, {Lee}, {Leitch}, {Lueker}, {Luong-van}, {Malkan}, {McIntyre},
  {McMahon}, {Mehl}, {Menten}, {Meyer}, {Mocanu}, {Murphy}, {Natoli}, {Padin},
  {Plagge}, {Reichardt}, {Rest}, {Ruel}, {Ruhl}, {Sharon}, {Schaffer}, {Shaw},
  {Shirokoff}, {Spilker}, {Stalder}, {Staniszewski}, {Stark}, {Story},
  {Vanderlinde}, {Welikala}, \& {Williamson}}]{vieira:13}
{Vieira}, J.~D., {Marrone}, D.~P., {Chapman}, S.~C., {et~al.} 2013,
  \href{http://dx.doi.org/10.1038/nature12001}{\JournalTitle{\nat}, 495, 344}

\bibitem[{{Wong} {et~al.}(2017){Wong}, {Ishida}, {Tamura}, {Suyu}, {Oguri}, \&
  {Matsushita}}]{Wong:17}
{Wong}, K.~C., {Ishida}, T., {Tamura}, Y., {et~al.} 2017,
  \href{http://dx.doi.org/10.3847/2041-8213/aa7d4a}{\JournalTitle{\apjl}, 843,
  L35}

\bibitem[{{Wong} {et~al.}(2015){Wong}, {Suyu}, \& {Matsushita}}]{Wong:15}
{Wong}, K.~C., {Suyu}, S.~H., \& {Matsushita}, S. 2015,
  \href{http://dx.doi.org/10.1088/0004-637X/811/2/115}{\JournalTitle{\apj},
  811, 115}

\end{thebibliography}

\end{document}